\documentclass[pra,aps,twocolumn,10pt,letterpaper,superscriptaddress]{revtex4-2}
\usepackage[utf8]{inputenc}
\usepackage[english]{babel}
\usepackage{bm}
\usepackage{graphicx}
\usepackage{amsmath}
\usepackage{amssymb}
\usepackage{amsfonts}
\usepackage{xcolor}
\usepackage{hyperref}
\hypersetup{
    pdfstartview={FitH},    % fits the width of the page to the window
    colorlinks=true,       % false: boxed links; true: colored links
    linkcolor=blue,          % color of internal links
    citecolor=blue,        % color of links to bibliography
    filecolor=blue,      % color of file links
    urlcolor=blue           % color of external links
}
\usepackage{multirow}

\usepackage{color}
\definecolor{BLUE}{rgb}{0.0,0.0,1.0}
\newcommand{\rn}[1]{{\color{BLUE}{#1}}}
\usepackage{soul}
\soulregister\cite7
\soulregister\ref7

\begin{document}
%
%%%%%%%%%%%%%%%%%%%%%%%%%%%%%%%%%%%%%%%%%%%%%%%%%%%%%%%%%%%%%%%%%%%%%%
\title{Orbital collapse and dual states of the $5g$ electrons \\
in superheavy elements}
%%%%%%%%%%%%%%%%%%%%%%%%%%%%%%%%%%%%%%%%%%%%%%%%%%%%%%%%%%%%%%%%%%%%%%
\author{I. I. Tupitsyn}
\email{i.tupitsyn@spbu.ru}
\affiliation{Department of Physics, St.~Petersburg State University, Universitetskaya 7-9, St.~Petersburg 199034, Russia}
\author{I. M. Savelyev}
\affiliation{Department of Physics, St.~Petersburg State University, Universitetskaya 7-9, St.~Petersburg 199034, Russia}
\author{Y. S. Kozhedub}
\affiliation{Department of Physics, St.~Petersburg State University, Universitetskaya 7-9, St.~Petersburg 199034, Russia}
\author{D. A. Telnov}
\affiliation{Department of Physics, St.~Petersburg State University, Universitetskaya 7-9, St.~Petersburg 199034, Russia}
\author{N.~K.~Dulaev}
\affiliation{Department of Physics, St.~Petersburg State University, Universitetskaya 7-9, St.~Petersburg 199034, Russia}
\affiliation{Petersburg Nuclear Physics Institute named by B.~P.~Konstantinov of National Research Center ``Kurchatov Institute'', Orlova roscha 1, Gatchina 188300, Leningrad region, Russia}
\author{A. V. Malyshev}
\affiliation{Department of Physics, St.~Petersburg State University, Universitetskaya 7-9, St.~Petersburg 199034, Russia}
\affiliation{Petersburg Nuclear Physics Institute named by B.~P.~Konstantinov of National Research Center ``Kurchatov Institute'', Orlova roscha 1, Gatchina 188300, Leningrad region, Russia}
\author{E. A. Prokhorchuk}
\affiliation{Department of Physics, St.~Petersburg State University, Universitetskaya 7-9, St.~Petersburg 199034, Russia}
\author{V. M. Shabaev}
\affiliation{Department of Physics, St.~Petersburg State University, Universitetskaya 7-9, St.~Petersburg 199034, Russia}
\affiliation{Petersburg Nuclear Physics Institute named by B.~P.~Konstantinov of National Research Center ``Kurchatov Institute'', Orlova roscha 1, Gatchina 188300, Leningrad region, Russia}
%
%%%%%%%%%%%%%%%%%%%%%%%%%%%%%%%%%%%%%%%%%%%%%%%%%%%%%%%%%%%%%%%%%%%%%%
\begin{abstract}
The problem of orbital collapse of the $5g$ and $6f$ electrons
in  atoms of superheavy elements (SHE) is considered. Previously, the 
presence of the orbital collapse was established for the $4f$ and $5f$ elements of
the periodic table. Because of the large centrifugal term for the $f$ and $g$
electrons, the effective radial potential has two wells, one narrow and deep
and the other wide but shallow. Depending on the external parameters,
the electron can be either localized in the outer well with low binding energy
and large average radius or in the inner one with higher energy and smaller radius.
In this work, we demonstrate the existence of the orbital collapse for the
$5g$ electrons when changing the total angular momentum $J$ of the atom.
We also found that for some SHE elements, two different solutions of the same Dirac-Fock equations may coexist, with the $5g$ electron localized either in the inner or outer well. In both cases, the radial wave functions are nodeless. The problem of the dual-state coexistence is studied
by the configuration-interaction method in the Dirac-Fock-Sturm orbital basis as well.
\end{abstract}
\maketitle
%%%%%%%%%%%%%%%%%%%%%%%%%%%%%%%%%%%%%%%%%%%%%%%%%%%%%%
\section{Introduction}
%%%%%%%%%%%%%%%%%%%%%%%%%%%%%%%%%%%%%%%%%%%%%%%%%%%%%%
The orbital-collapse phenomenon was first predicted in
Refs.~\cite{Fermi_1928,Mayer_1941}. It was shown that due to
the large size of the repulsive centrifugal term, the effective radial potential
acting on the $4f$ and $5f$ electrons can have two potential wells: a deep and
narrow inner well and a shallow but wide outer well.
The formation of the double-well radial potential
is determined by the magnitude of the centrifugal term, which
increases quadratically
with the growth of the orbital quantum number~$l$. Depending on the external
parameters, the $f$~orbital can be localized either in the inner well
or in the outer one. When these parameters change, an electron initially localized,
for example, in the external well, can move into the internal well.
At the same time, the radius of the $f$ orbital sharply decreases
tenfold, which can lead to a sudden change in various physical and chemical
properties of the atom.

As shown in Refs.~\cite{Cowan_1968, Griffen_1969}, an orbital
collapse of the $d$ electrons can also take place for the excited
states of atoms. In Ref.~\cite{Griffen_1969}, the possibility
of the collapse of the $g$ electrons in superheavy elements (SHE) was predicted as well.
Orbital collapse can occur in the isoelectronic sequence of atoms and ions~\cite{Cowan_1968, Cheng_1983, Migdalek_1984, Migdalek_2000}, in a series
of sequentially ionized atoms~\cite{Lucatorto_1981}, in confined 
(in cavity) and compressed atoms~\cite{Connerade_2000, Connerade_2020},
in the homologous sequence of the periodic 
table~\cite{Connerade_1991}, as a function of the chemical environment of the atom 
\cite{Maiste_1980, Ruus_1999}, in a series of different atomic terms
of the same configuration \cite{Tupitsyn_2023}, and so on.
The effect of the orbital collapse can manifest itself in various experiments, e.g., in photoabsorption, photoionization, etc.
\textit{Ab initio} calculations devoted to the study of the orbital\rn{-}collapse problem
were performed previously both by the nonrelativistic Hartree-Fock 
~\cite{Connerade_1978, Karaziya_1981, Cheng_1983} and 
relativistic Dirac-Fock (DF) \cite{ Band_1980, Band_1981, Migdalek_1984,
Migdalek_1987, Migdalek_2000, Tupitsyn_2023} methods. It should be noted
that the central-field approximation, employed in the non-relativistic and
relativistic versions of the Hartree-Fock method, does not allow one to study
the orbital collapse-effect for atomic configurations involving more than
one electron in the shell of interest, (for  example, $4f$ shell).
The reason is that in the central-field approximation all electrons in the shell possess the same radial wave function. 
As a result, the state of the atom with more than one electron in the outer well is certainly energetically inefficient.

As already noted, the orbital collapse usually occurs when some external parameters change.
As a consequence, an electron can move from the external well to 
the internal one.
However, in Ref.~\cite{Band_1980} it was found that within the
framework of the DF method
two different solutions of the same self-consistent field (SCF)
equations can be obtained without changing any external parameters.
In one of them, the $4f$ electron is localized in the inner well, whereas in the other it is localized in the outer well.
 In Ref.~\cite{Band_1980}, the coexistence of two different solutions with the same atomic configuration, the ``blow''
and the ``collapse'' ones, was shown for the excited state of lanthanum ([Xe]$6s^24f_{5/2}$) and the ground state of europium ([Xe]$6s^24f^6_{5/2}4f_{7/2}$).
It is noteworthy that both the ``blow'' and ``collapse'' $4f$ orbitals are nodeless. They can be obtained as the solutions of the SCF equations, provided the initial approximation is appropriately chosen, and correspond to two different stationary values of the DF energy functional. Thus, the DF operator in the converged SCF equations is also different for these two solutions. That is why the coexistence of two different nodeless orbitals with the same quantum numbers does not contradict the Sturm's oscillation and separation theorems
\cite{Atkinson_1964}.

In this work, we investigate the problem of the $5g$-electron collapse in atoms
of the eighth-period SHEs. As was shown in our work
\cite{Savelyev_2023} and in the papers~\cite{Fricke_1977, Nefedov_2006, 
Smits_2023a, Smits_2023b},
the occupation of the $5g$~shell in the ground state starts at $Z=125$ and continues up to $Z=144$
(from the multiconfiguration calculations, it follows that this shell becomes closed at $Z=145$). 
Since the DF method allows us to study the collapse for only one
electron on the $5g_{7/2}$~shell, we restrict  ourselves to the
calculations of the ground configuration 
[Og]$8s^2 8p_{1/2}^1 6f_{5/2}^3 5g_{7/2}^1$
of the atom with $Z = 125$ as well as the excited configurations
[Og]$8s^2 8p_{1/2}^1 6f_{5/2}^2 5g_{7/2}^1$ and
[Og]$8s^2 8p_{1/2}^2 6f_{5/2}^3 5g_{7/2}^8  5g_{9/2}^1 $ for $Z=124$ and $Z=134$,
respectively. We also obtain the $6f$-orbital dual solutions for the configuration 
[Og]$8s^2 8p_{1/2}^2 5g^{18} 7d^1_{3/2} 6f_{5/2}^6 6f_{7/2}^1$
of the SHE with $Z=148$.
The calculations are performed by the single-configuration DF
method \cite{Bratzev_1977} for individual atomic terms with the given total
angular momentum~$J$ as well as in the approximation of the gravity-center of the relativistic configuration \cite{Grant_1970, Tupitsyn_2018}.
The dual solutions of the DF equations for the aforementioned elements and configurations are obtained using the different initial approximations. In addition,
we reproduce the obtained in Ref. \cite{Band_1980} dual solutions, the ``blow'' and the ``collapse'' ones,
for lanthanum ([Xe]$6s^24f_{5/2})$ and europium ([Xe]$6s^24f^6_{5/2}4f_{7/2}$).
Finally, in order to determine the mixing
of the different many-electron states with the localized and delocalized $5g$ orbitals, we perform the configuration-interaction (CI) calculations in the basis of the Dirac-Fock-Sturm (DFS) orbitals 
\cite{Tupitsyn_2003A, Tupitsyn_2003B}.

Atomic units (a.u.) are used throughout the paper unless explicitly
stated otherwise.
%
%%%%%%%%%%%%%%%%%%%%%%%%%%%%%%%%%%%%%%%%%%%%%%%%%%%%%%
\section{Details of the calculations}\label{sec:details}
%%%%%%%%%%%%%%%%%%%%%%%%%%%%%%%%%%%%%%%%%%%%%%%%%%%%%%
%
In calculations of the DF one-electron wave functions and energies, we use the many-electron Dirac-Coulomb Hamiltonian $\hat H^{\rm DC}$: 
\begin{align}\label{eq:H^DC}
\hat{H}^{\rm DC} = \hat{H}^{\rm D} + \hat{V}^{\rm C} ,
\end{align}
where~$\hat{H}^{\mathrm{D}}$ is the sum of the one-electron Dirac
Hamiltonians,
\begin{equation}
\hat{H}^{\mathrm{D}} = \sum_{i=1}^N \left[ (\bm{\alpha}_i \cdot \bm{p}_i)c+
(\beta_i-1)mc^2+V_{\rm n}(r_i) \right].
\end{equation}
Here $\bm{\alpha}$ is a vector of the Dirac matrices and
$\hat{V}^{\rm C}$ is the sum of the Coulomb electron-electron interaction operators,
\begin{equation}
    \hat{V}^{\rm C} = \frac{1}{2}\sum_{i\neq j}^N \frac{1}{r_{ij}}.
\end{equation}
All the calculations are performed with the nuclear potential $V_{\rm n}(r)$ constructed employing the Fermi model for the nuclear-charge distribution.
The root-mean-square (RMS) radius of the SHE nucleus (in fm) is given by 
\begin{equation}
R = \sqrt{\frac{3}{5}}R_{\mathrm{sphere}}, 
\qquad R_{\mathrm{sphere}} = 1.2 A^{1/3},
\end{equation}
where for the nucleon number $A$ we use the approximate formula from
Ref.~\cite{Pieper_1969}, 
\begin{equation}\label{eq:nucleon}
A = 0.00733 Z^2 + 1.30 Z + 63.6.
\end{equation}
The value of $A$ obtained from Eq.~(\ref{eq:nucleon}) is rounded to the
nearest integer. This choice of the nuclear size is consistent with the
one made in Ref.~\cite{Savelyev_2023}.
The RMS radii of \rn{the} La and Eu atoms are taken to be
4.8550~fm ($A=139$)  and 5.1115~fm ($A=153$), respectively 
\cite{Angeli_2013}.

In our DF calculations, the modified G{\'a}sp{\'a}r~\cite{Gaspar_1952} potential
$V_{\rm G}(r)$ is used as an initial approximation in the SCF
procedure. The modification is made by taking into account the
self-interaction correction (SIC)~\cite{Green_1973}. The employed potential reads as
\begin{equation}
V_{\rm G}(r)=-\frac{Z}{r}+\frac{N_e-1}{r} \,
\left(1-\frac{e^{- \lambda r}}{1+b\, r} \right ) \,,
\label{Gaspar}
\end{equation}
where $\lambda=0.2075 \, Z^{1/3}$,  $b=1.19 \, Z^{1/3}$, and $N_e$ is
the number of electrons. 

It should be noted that in all the cases the $5g$ orbital obtained
by solving the one-electron Dirac equation with the local G{\'a}sp{\'a}r potential
$V_{\rm G}(r)$ is localized in the outer well. In order to manage the convergence process in the SCF calculations, we introduce a real parameter $\alpha$ into the Dirac-Fock operator $V_{\rm DF}$
\begin{equation}
\label{eq:V_alpha}
V_{\rm DF}(\alpha,r)=V_{\rm H}(r)+ \alpha \, V_{\rm x},
\end{equation}
where $V_{\rm H}$ is the Hartree potential with the SIC and 
$V_{\rm x}$ is the exchange operator. By changing $\alpha$ from zero
to unity, the contribution of the exchange interaction can be controlled.
This contribution affects the localization of the $5g$ electron in either the inner or the outer well during the SCF calculations. Naturally, at the end of the SCF procedure, when the convergence is achieved, $\alpha$ must be equal to unity.

The single-configuration and CI total energies are calculated using the Dirac-Coulomb-Breit Hamiltonian $\hat H^{\rm DCB}$: 
\begin{align}\label{eq:H^DCB}
\hat{H}^{\rm DCB} =\Lambda^{+} 
\left( \hat{H}^{\rm DC} + \hat{V}^{\rm B} \right) \Lambda^{+} \,,
\end{align}
where $\Lambda^{+}$ is the product of the one-electron projectors on the
positive-energy solutions of the DF equations and $\hat{V}^{\rm B}$
is the Breit-interaction operator,
\begin{equation}
\hat{V}^{\rm B} = -\frac{1}{2} \sum_{i \neq j}^N \frac{1}{2r_{ij}}
\Big[\bm{\alpha}_i\cdot\bm{\alpha}_j+\frac{(\bm{\alpha}_i
\cdot\bm{r}_{ij})(\bm{\alpha}_j\cdot\bm{r}_{ij})}{r_{ij}^2}\Big].
\end{equation}
%
%%%%%%%%%%%%%%%%%%%%%%%%%%%%%%%%%%%%%%%%%%%%%%%%%%%%%%%%%%%%%%%%%%%%%
\section{Effective radial potential}\label{sec:ERP}
\begin{figure}
\includegraphics[width=\columnwidth]{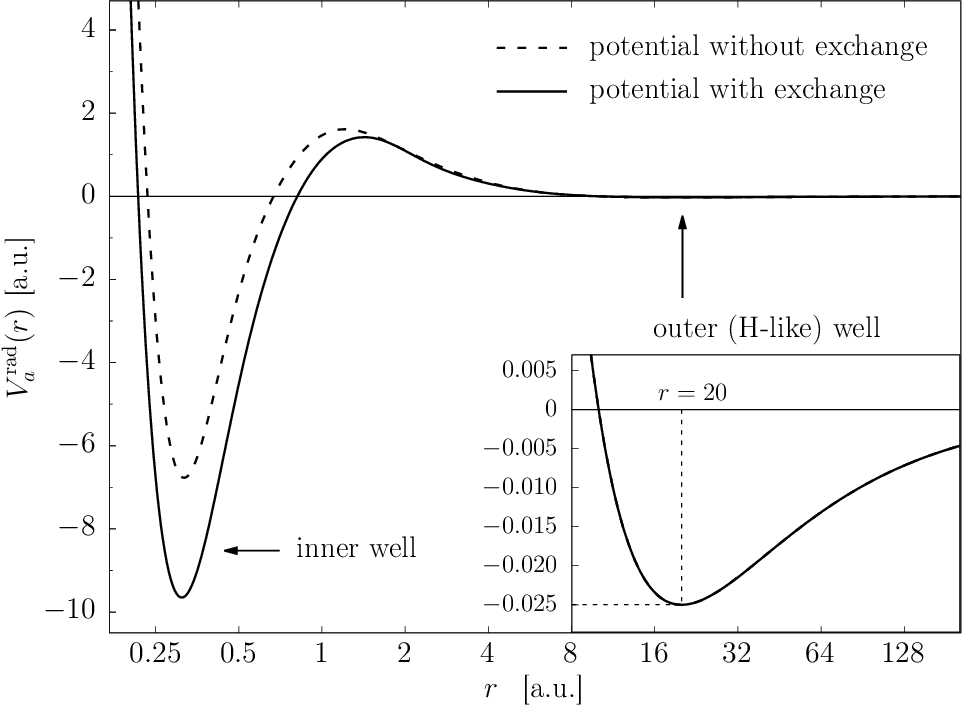}
\caption{\label{fig1} 
Effective radial potential $V_{a}^{\rm rad}(r)$ for
the shell $a$ = $5g_{7/2}$ of the superheavy atom with $Z = 125$. 
The solid line is the potential with the exchange, while the dashed line
is the potential without the exchange.}
\end{figure}

%%%%%%%%%%%%%%%%%%%%%%%%%%%%%%%%%%%%%%%%%%%%%%%%%%%%%%%%%%%%%%%%%%%%%
%
In the DF method, the one-electron radial potential $V^{\rm DF}_{a}$
for the shell $a$ is a nonlocal operator. For this reason, solely
to demonstrate the behavior of the effective radial potential, we replace
the nonlocal operator $V^{\rm DF}_{a}$ by the so-called local DF
potential~$V^{\rm loc}_{a}(r)$~\cite{Shabaev_2005}. 
An example of the effective radial potential $V^{\rm rad}_a(r)$ obtained
for the SHE with $Z=125$ with and without taking into account the exchange is shown in Fig.~\ref{fig1}.
This potential has two wells (for convenience, the outer well is shown
in an enlarged scale in the lower right corner).
As can be seen from Fig.~\ref{fig1}, the exchange potential actually affects only the depth
of the inner well, as it decreases exponentially with increasing the distance.

In the nonrelativistic 
approximation the effective radial potential $V^{\rm rad}_a(r)$ can be
represented as the sum of the local potential $V^{\rm loc}_{a}(r)$ and
the centrifugal term
\begin{equation}
V^{\rm rad}_a(r)= V^{\rm loc}_{a}(r)+ \frac{l_a(l_a+1)}{2r^2} \,,
\label{pot1}
\end{equation}
where $l_a$ is the orbital quantum number.
Because of the rapid exponential decay of the exchange term, the asymptotic form of the local potential~$V^{\rm loc}_a(r)$ at large
distances is purely Coulombic. Thus, the radial potential~$V_{a}^{\rm rad}(r)$
for a neutral atom in the asymptotic region has the form
\begin{equation}
V^{\rm rad}_a(r) \to -\frac{1}{r}+ \frac{l_a(l_a+1)}{2r^2} \,, \qquad
r \to \infty \,.
\label{pot2}
\end{equation}
For this reason, for large $l_a$, the position of the local minimum $r^{\rm min}_a$ of $V_{a}^{\rm rad}(r)$ corresponding to the outer well and its depth $V^{\rm min}_a$ can be determined with a high degree of accuracy using the expressions:
\begin{equation}
r^{\rm min}_{a}=l_a(l_a+1) \,, \qquad 
V^{\rm min}_{a}=\frac{1}{2l_a(l_a+1)} \,. 
\label{pot3}
\end{equation}
These results are supported by Fig.~~\ref{fig1}, where for the $g$ orbital one should set $l_a=4$. 
Indeed, the minimum of the shallow and wide outer well is located at $r = 20$~a.u., while the depth is of the order of 0.025 a.u. The deep and narrow inner well has the minimum at a small distance $r = 0.32$~a.u.

We stress also that in the case of a neutral atom the electron in the outer well can be
considered as an electron in the field of a singly charged ion, whose
potential at large distances approximately coincides with the Coulomb
potential $V_{a}(r)=-1/r$. Thus, the one-electron energy and the mean radius
of the electron in the outer well must be close to the energy 
$\varepsilon^{\rm H}_{a}$ and the mean radius
$\langle r \rangle^{\rm H}_{a}$ for the hydrogen atom:
\begin{equation}
\label{eq:H-like}
\varepsilon^{\rm H}_{a}=-\frac{1}{2n_a^2}\,, \qquad 
\langle r \rangle^{\rm H}_{a} = \frac{1}{2} \, 
\left[3n_a^2 -l_a(l_a+1) \right] \,.
\end{equation} 
where $n_a$ is the principal quantum number.
In the case of the $5g$ electron\rn{:}  $\varepsilon^{\rm H}_{5g}= -0.02$ a.u.
and $\langle r \rangle^{\rm H}_{5g} =27.5$~a.u. 
This statement is confirmed by the direct calculations below.
% 
%%%%%%%%%%%%%%%%%%%%%%%%%%%%%%%%%%%%%%%%%%%%%%%%%%%%%%%%%%%%%%%%%%%%%
\section{\label{sec:orbital_collapse} Orbital collapse}
%%%%%%%%%%%%%%%%%%%%%%%%%%%%%%%%%%%%%%%%%%%%%%%%%%%%%%%%%%%%%%%%%%%%%
%
In this section, we discuss the orbital collapse of the $5g$ electron
in the course of changing the value of the total angular momentum
$J$ on the examples of the ground configuration for $Z=124$ 
([Og]$8s^28p_{1/2}^16f_{5/2}^3 5g_{7/2}^1$)
and the excited configuration for $Z=125$ 
([Og]$8s^28p_{1/2}^16f_{5/2}^2 5g_{7/2}^1$).
In Tables \ref{tab_Z125} and \ref{tab_Z124}, for each value of the total 
angular momentum $J$, the values of the one-electron energies 
and average radii for the $5g_{7/2}$ shell as well as the total energies of atoms 
are given for $Z=125$ and $Z=124$, respectively.

\begin{table}
\caption{\label{tab_Z125} 
$Z=125$ ([Og]$8s^2 8p_{1/2}^1 6f_{5/2}^3 5g_{7/2}^1$).
One-electron energies $\varepsilon_{5g}$ and average radii 
$\langle r\rangle_{5g}$ of the valence $5g_{7/2}$ orbital
and the total energies $E_{\rm DCB}$ of the neutral atom.
$K$ is the number of energy levels with the given $J$ in the configuration. 
The energies are shown with the opposite sign.
All the values are given in atomic units.}
%%%%%%%%%%%%%%%%%%%%%%%%%%%%%%%%%%%%%%%%%%%%%%%%%%%%%%%%%%%%%%%%%%%%%%%%%
\begin{ruledtabular}
\begin{tabular}{cccrc}
$J$ & $K$ &  $-\varepsilon_{5g}$  & $\langle r\rangle_{5g}$  & 
$-E_{\rm DCB}$ \\
\hline
1/2  & 2 &  $ 0.0200016$ &  27.494 &  $ 64718.58334$ \\
3/2  & 5 &  $ 0.0200015$ &  27.494 &  $ 64718.59179$ \\
5/2  & 6 &  $ 0.0200017$ &  27.493 &  $ 64718.59179$ \\
7/2  & 6 &  $ 0.0200017$ &  27.493 &  $ 64718.59179$ \\
9/2  & 6 &  $ 0.0200017$ &  27.493 &  $ 64718.59179$ \\
11/2 & 5 &  $ 0.0200019$ &  27.493 &  $ 64718.59179$ \\
13/2 & 3 &  $ 0.5387971$ &   0.732 &  $ 64718.85035$ \\
15/2 & 2 &  $ 0.5348849$ &   0.732 &  $ 64718.84021$ \\
17/2 & 1 &  $ 0.5367741$ &   0.733 &  $ 64718.84636$ \\
\end{tabular}
\end{ruledtabular}
\end{table}
 
\begin{table}
\caption{\label{tab_Z124} 
$Z=124$ ([Og]$8s^2 8p_{1/2}^1 6f_{5/2}^2 5g_{7/2}^1$).
One-electron energies $\varepsilon_{5g}$ and average radii 
$\langle r\rangle_{5g}$ of the valence $5g_{7/2}$ orbital
and the total energies $E_{\rm DCB}$ of the neutral atom.
The notations are the same as in Table~\ref{tab_Z125}.
All the values are given in atomic units. }
%%%%%%%%%%%%%%%%%%%%%%%%%%%%%%%%%%%%%%%%%%%%%%%%%%%%%%%%%%%%%%%%%%%%%%%%%
\begin{ruledtabular}
\begin{tabular}{cccrc}
$J$ & $K$ & $-\varepsilon_{5g}$  & $\langle r\rangle_{5g}$ &
$-E_{\rm DCB}$ \\
\hline
  0    &  1  &     $ 0.01996061$   &    27.567    &    $ 63308.54698$  \\ 
  1    &  3  &     $ 0.01998763$   &    27.520    &    $ 63308.55467$  \\
  2    &  4  &     $ 0.01999968$   &    27.497    &    $ 63308.55467$  \\
  3    &  5  &     $ 0.01999626$   &    27.504    &    $ 63308.55460$  \\
  4    &  5  &     $ 0.01999806$   &    27.499    &    $ 63308.55462$  \\
  5    &  4  &     $ 0.02001107$   &    27.475    &    $ 63308.55472$  \\
  6    &  3  &     $ 0.02002020$   &    27.457    &    $ 63308.55474$  \\
  7    &  2  &     $ 0.24072513$   &     0.799    &    $ 63308.52478$  \\
  8    &  1  &     $ 0.23380541$   &     0.799    &    $ 63308.50880$  \\
\end{tabular}
\end{ruledtabular}
\end{table}

It can be seen from Table~\ref{tab_Z125} that for all the values of $J$
from $J=1/2$ up to $J=11/2$  the $5g$ electron has a very large radius
and, hence, it is localized in the outer well. The values of the one-electron energies
and mean radii are very close to the hydrogen values, see the discussion at the end of the previous section. For $Z=125$, the orbital
collapse occurs at the transition from $J=11/2$ to $J=13/2$.
As a result,
the one-electron energy  $\varepsilon^{\rm H}_{5g}$ increases more than
$25$ times in magnitude, the mean radius $\langle r \rangle^{\rm H}_{5g}$ decreases
almost $40$ times, and the total energy changes by about $0.25$~a.u.

A similar effect is observed for $Z =124$, as demonstrated in Table~\ref{tab_Z124}.
In this case, the orbital collapse of the $5g$ orbital occurs at $J \geqslant 7$. 
However, in contrast to $Z =125$, in this case the total energies of the atomic terms
with $J=7,8$ are lower by $0.03$ and $0.05$~a.u. than the energies of the terms $J \leqslant 6$, for which the $5g$ orbital is localized in the outer well.
We note that in Table~\ref{tab_Z125} the total energies for all $J$ in the range $3/2\leqslant J \leqslant 11/2$
 almost coincide with each other and equal $-64718.59179$~a.u.
This value is very close to the value
$-64718.59196$, which is 
obtained by summing
the total energy of the lowest-energy state of the ion with the $5g$ electron removed ($E^{\rm ion}=-64718.57196$) and the one-electron energy of the $5g_{7/2}$ orbital ($\varepsilon_{5g}=-0.020002$); the corresponding ion has $J_{\rm ion} = 5$.
Since a polarization of the ionic electron shells by the $5g$ electron located
at a very large distance is very weak, all the states of the neutral atom with the total angular momentum
$|J_{\rm ion} - 7/2| \leqslant J \leqslant |J_{\rm ion} + 7/2|$ have practically the same energy.

\begin{figure}
\includegraphics[width=\columnwidth]{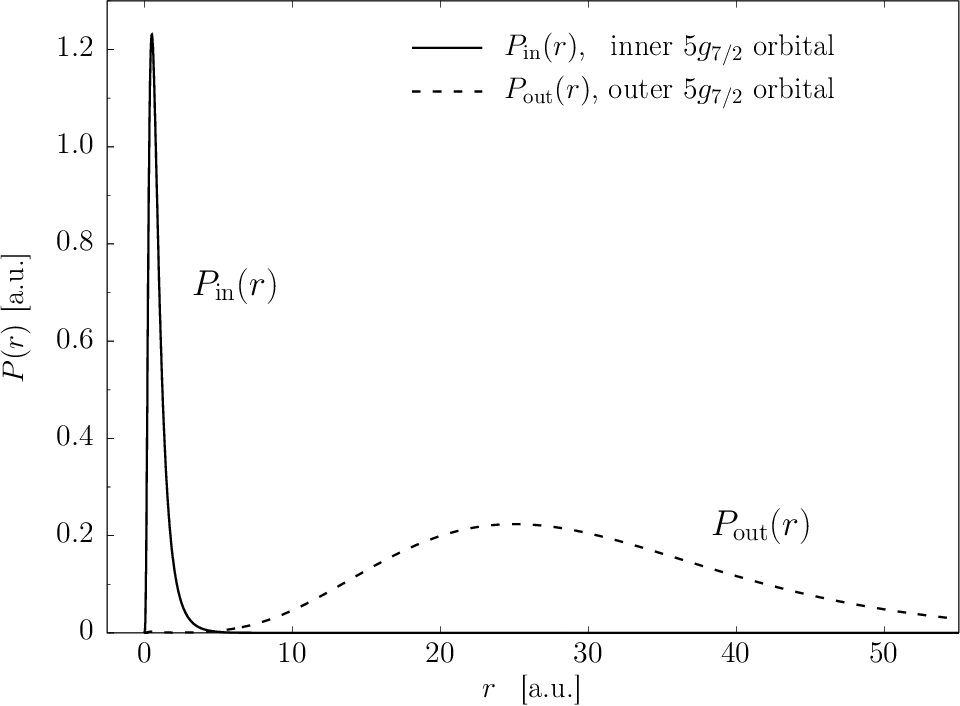}\\
\caption{\label{Fig_Z125} 
$Z=125$ ([Og]$8s^2 8p_{1/2}^1 6f_{5/2}^3 5g_{7/2}^1$).
Large components of the radial wave functions: 
solid line ($P_{\rm in}(r)$) and dashed line ($P_{\rm out}(r)$) correspond to the  $5g_{7/2}$ orbitals
localized in the inner and outer wells, respectively.
All the values are given in atomic units.}
\end{figure}
Fig.~\ref{Fig_Z125} presents the large components of two radial 
$5g_{7/2}$ wave functions. One of them, $P_{\rm in}(r)$, is localized
in the inner well (the solid line) and the other one, $P_{\rm out}(r)$, is localized in the
outer well (the dashed line). Despite the fact that both functions are nodeless,
they overlap very weakly. The overlap integral 
$ \langle P_{\rm out}\mid P_{\rm in} \rangle $ is of the order of $0.001$.
%
%%%%%%%%%%%%%%%%%%%%%%%%%%%%%%%%%%%%%%%%%%%%%%%%%%%%%%%%%%%%%%%%%%%%%
\section{\label{sec:dual_states} Dual states}
%%%%%%%%%%%%%%%%%%%%%%%%%%%%%%%%%%%%%%%%%%%%%%%%%%%%%%%%%%%%%%%%%%%%%
%
The dual solutions of the
DF equations, the ``blow'' and the ``collapse'' ones, were first obtained in lanthanum ($Z=57$, [Xe]$6s^24f^1_{5/2}$) and 
europium ($Z=63$, [Xe]$6s^24f^6_{5/2}4f^1_{7/2}$) in Ref. \cite{Band_1980}.
It was done by ``pushing'' the $4f$ electron at the first iterations slightly
to the inner or outer wells (see the details in
Ref.~\cite{Band_1980}).
In the present work, we have also found the dual solutions in La and Eu. We calculated the total energies both with ($E_{\rm DCB}$)
and without ($E_{\rm DC}$) the Breit interaction in order to compare
our data with the results of Ref.~\cite{Band_1980}, where the Breit
interaction was not taken into account.

To obtain the $4f$ orbitals localized in the inner well, we have included the full exchange interaction, i.e., $\alpha=1$ in Eq.~(\ref{eq:V_alpha}). 
The $4f$ orbital localized in the outer well was obtained by the following
way. At the first stage, we have performed the SCF calculation excluding the exchange
interaction, i.e., $\alpha=0$. In the no-exchange case, the inner well becomes
less deep (see Fig.~\ref{fig1}), and the solution of the Hartree equation
with the SIC for the $4f$ electron localizes in the outer well. On the next
step, we have repeated the SCF calculation with a full exchange, $\alpha=1$,
using the one-electron functions obtained at the first stage as the initial
approximation.

\begin{table}
\caption{\label{tab_La} 
La, $Z=57$ ([Xe]$6s^2 4f_{5/2}^1$).
One-electron energies $\varepsilon_{4f}$ and average radii 
$\langle r\rangle_{4f}$ of the valence $4f_{5/2}$ orbital
and the total energies of the neutral atom with ($E_{\rm DCB}$) and without ($E_{\rm DC}$) the Breit interaction. The labels ``in'' and ``out''
indicate two dual states localized in the inner and outer wells, respectively. 
The column ``Hydrogen'' shows the values obtained according to Eq.~(\ref{eq:H-like}). 
All the values are given in atomic units.}
%%%%%%%%%%%%%%%%%%%%%%%%%%%%%%%%%%%%%%%%%%%%%%%%%%%%%%%%%%%%%%%%%%%%%%%%%
\begin{ruledtabular}
\begin{tabular}{cccc}
&  Ref. \cite{Band_1980}  & This work & Hydrogen \\
\hline
$-\varepsilon_{4f}^{\rm in}$ &  0.2381     &  0.23830    &      --  \\
$-\varepsilon_{4f}^{\rm out}$ &  0.0316     &  0.03180    &   0.03125 \\
$\langle r\rangle_{4f}^{\rm in}$          &  ~~~1.2591     &  ~~1.2591     &      --  \\
$\langle r\rangle_{4f}^{\rm out}$          &  17.062     &  17.0614    &   18.0000 \\
$-E_{\rm DC}^{\rm in}$           &  8493.6246  &  8493.5521  &      --  \\
$-E_{\rm DC}^{\rm out}$           &  8493.5512  &  8493.4767  &      --  \\
$-E_{\rm DCB}^{\rm in}$           &    --       &  8486.5883  &      --  \\
$-E_{\rm DCB}^{\rm out}$          &    --       &  8486.5096  &      --  \\
\end{tabular}
\end{ruledtabular}
\end{table}
\begin{table}
\caption{\label{tab_Eu} 
Eu, $Z=63$ ([Xe]$6s^2 4f_{5/2}^6 4f_{7/2}^1$).
One-electron energies $\varepsilon_{ 4f}$ and average radii 
$\langle r\rangle_{4f}$ of the valence $4f_{7/2}$ orbital
and the total energies of the neutral atom with ($E_{\rm DCB}$) and without ($E_{\rm DC}$) the Breit interaction.
The notations are the same as in Table~\ref{tab_La}. All the values are given in atomic units.}
%%%%%%%%%%%%%%%%%%%%%%%%%%%%%%%%%%%%%%%%%%%%%%%%%%%%%%%%%%%%%%%%%%%%%%%%%
\begin{ruledtabular}
\begin{tabular}{cccc}
&  Ref. \cite{Band_1980}  & This work & Hydrogen \\
\hline
$-\varepsilon_{4f}^{\rm in}$  & 0.3609     & 0.36073    &      --  \\
$-\varepsilon_{ 4f}^{\rm out}$  & 0.0316     & 0.03146    &  0.03125 \\
$\langle r\rangle_{ 4f}^{\rm in}$          &  ~~~0.94877    &  ~~0.94877   &      --  \\
$\langle r\rangle_{4f}^{\rm out}$          &  17.709     &  17.7087   &   18.0000 \\
$-E_{\rm DC}^{\rm in}$           & 10846.7756  & 10846.6533 &      --  \\
$-E_{\rm DC}^{\rm out}$           & 10846.6727  & 10846.5483 &      --  \\
$-E_{\rm DCB}^{\rm in}$          &     --      & 10836.8226 &      --  \\
$-E_{\rm DCB}^{\rm out}$          &     --      & 10836.7123 &      --  \\
% \bottomrule
\hline
\end{tabular}
\end{ruledtabular}
\end{table}

\begin{table*}
\caption{\label{tab_SHE} Dual states of the SHE atoms with $Z=124, 125, 134, 148$.
One-electron energies $\varepsilon_{a}$ 
and average radii  $\langle r \rangle_{a}$ of the collapsing $a$ orbitals and the total energies $E_{\rm DCB}$ of the neutral atoms.
The labels ``in'' and ``out'' indicate two dual states localized in the inner and outer wells, respectively. All the values are given in atomic units. }
%%%%%%%%%%%%%%%%%%%%%%%%%%%%%%%%%%%%%%%%%%%%%%%%%%%%%%%%%%%%%%%%%%%%%%%%%
\begin{ruledtabular}
\begin{tabular}{ccccc}
 \multirow{ 2}{*}{Property} & {$Z=124$\footnotemark[1]} & {$Z=125$\footnotemark[2]} & {$Z=134$\footnotemark[3]} & 
{$Z=148$\footnotemark[4]} \\[-1mm]
                            & {$a=5g_{7/2},~J=7$} & {$a=5g_{7/2},~J=6.5$} & 
{$a=5g_{9/2},~J=6$} &  {$a=6f_{7/2},~J=4$}  \\
% \midrule
\hline

$-\varepsilon_{a}^{\rm in}$   & 0.240725  & 0.538797 & 0.534678  & 0.124697 \\

$-\varepsilon_{a}^{\rm out}$   & 0.019998  & 0.020002 & 0.020001  & 0.031689 \\

$\langle r \rangle_{a}^{\rm in}$ & 0.7990  & 0.7317  &  0.6304  &  1.5742  \\

$\langle r \rangle_{a}^{\rm out}$ & 27.5007 & 27.4934  &  27.4977  & 17.5938 \\

$-E_{\rm DCB}^{\rm in}$  & 63185.579  & 64718.850  & 80420.248  & 114885.579 \\
$-E_{\rm DCB}^{\rm out}$  & 63185.587  & 64718.592  & 80420.067  & 114885.578 \\
%%%%%%%%%%%%%%%%%%%%%%%%%%%%%%%%%%%%%%%%%%%%%%%%%%%%%%%%%%%%%%%%%%%%%%%%%
\end{tabular}
\end{ruledtabular}
\footnotetext[1]{[Og]$8s^2 8p_{1/2}^1 6f_{5/2}^2 5g_{7/2}^1$}
\footnotetext[2]{[Og]$8s^2 8p_{1/2}^1 6f_{5/2}^3 5g_{7/2}^1$}
\footnotetext[3]{[Og]$8s^2 8p_{1/2}^2 6f_{5/2}^3 5g_{7/2}^8  5g_{9/2}^1$}     
\footnotetext[4]{[Og]$8s^2 8p_{1/2}^2 5g^{18} 7d^1_{3/2} 6f_{5/2}^6 6f_{7/2}^1$}
\end{table*}
The results of the calculations of the dual states in La and Eu are presented in
Tables~\ref{tab_La} and \ref{tab_Eu}, respectively. As can be seen from
both tables, our data are in very good agreement with the results of Ref.
\cite{Band_1980}.  The one-electron energies $\varepsilon_{4f}$ of 
the $4f$ electron localized in the outer well are about 10 times smaller than the corresponding energies for
the inner well, whereas the average radii $\langle r\rangle_{4f}$ of the state
in the outer well are about $15$ times larger than their inner-well counterparts. It should be noted that the
one-electron energies and average radii of the $4f$ orbitals in the outer
well are close to the hydrogen values shown in the last columns in Tables~\ref{tab_La} and \ref{tab_Eu}. The comparison of the total energies
$E_{\rm DC}$ and $E_{\rm DCB}$ of the dual states shows that in both
cases the state with the electron localized in the inner well is
energetically more favorable.

The dual solutions for the atoms with $Z=124, 125$ (the corresponding configurations are indicated at the
bottom of Table~\ref{tab_SHE}) have been obtained as follows. As can be seen
from Table~\ref{tab_Z125}  and Table~\ref{tab_Z124}, when the G{\'a}sp{\'a}r potential
is used as the initial approximation, the collapse occurs in the transitions
 from the atomic state with total angular momentum $J=6$
to $J=7$ and from $J=11/2$ to $J=13/2$ for $Z=124$ and $Z=125$, respectively. In order to obtain the
solutions localized in the inner wells, it is sufficient to take the
orbitals obtained for the terms $J=7$ and $J=13/2$ as the initial approximations
for all the other values of $J$ for $Z=124$ and $Z=125$, respectively. On the
contrary, to obtain the orbitals localized in the outer wells, the orbitals
evaluated for $J=6$ and $J=11/2$ can be used as the initial approximations
for $Z=124$ and $Z=125$, respectively. For the SHE atoms with $Z=134$
and $Z=148$, also presented in Table~\ref{tab_SHE}, we have used the procedure described above for the Eu and La atoms.

The second, third, and fourth columns in Table~\ref{tab_SHE} present the one-electron
energies $\varepsilon_{5g}$ and the average radii $\langle r\rangle_{5g}$ of the $5g$ orbitals for two different nodeless radial solutions of the same
DF equations for the SHE atoms with $Z=124, 125, 134$. The fifth column
shows the similar results for two $6f$ orbitals localized in the inner and
outer wells for the element with $Z=148$. Both radial parts of these $6f$ orbitals
have two nodes. The notations ``in'' or ``out'' mean that the corresponding orbital is localized in the inner or in the outer well, respectively. 
The data reported in Table~\ref{tab_SHE} are  obtained for the atomic terms $J$, which have the lowest total energies
$E_{\rm DCB}$ for the configurations shown at the
bottom of the table.

It can be seen from Table~\ref{tab_SHE} that the one-electron energies and
the average radii of the $5g$ electrons in the outer well are very close
to the hydrogen values  $\varepsilon^{\rm H}_{\rm 5g}=0.02$~a.u. 
and  $\langle r\rangle^{\rm H}_{5g}=27.5$~a.u., respectively.
It is noteworthy that, although the localized in the outer well $6f$ orbital of the atom with $Z=148$ has two nodes at the points $r_1=0.193$~a.u.
and $r_2=0.510$~a.u., it is very similar to the nodeless hydrogen
$4f$ orbital at the larger distances. This is in consistency with the fact
that the values of the one-electron energy $0.03169$~a.u. and average radius
$17.5938$~a.u. for the $6f$ orbital in the outer well are close to the hydrogen values $\varepsilon^{\rm H}_{4f}=0.03125$~a.u. and
$\langle r\rangle^{\rm H}_{4f}=18.0$~a.u.

The total SHE energies given in the last two rows of Table~\ref{tab_SHE} are calculated by
diagonalizing the matrix of the DCB Hamiltonian 
(\ref{eq:H^DCB}) in the basis of the CSF
which are the eigenstates of the $\hat J^2$ and $\hat J_z$ operators.
The CSF are the linear combinations of the Slater determinants for a 
single relativistic configuration.
This approach is equivalent, in fact, to the single-configuration DF method.
The diagonalization of the Hamiltonian matrix is necessary because there are
many states with the given value of the total angular momentum $J$ in
the complex atomic configurations of the SHE (see Tables~\ref{tab_Z125}
and \ref{tab_Z124}).

In the present work we have obtained two different solutions of the DF equations
for the same configuration. An interesting question is how strongly they
interact with each other. For this purpose, we have performed the
calculations of the total energies and the many-electron wave functions for
the [Xe]$6s^2 4f_{5/2}$ configuration of the La atom by the CI method
in the DFS basis \cite{Tupitsyn_2003A, Tupitsyn_2003B}. 
The conventional DFS basis contains the one-electron DF functions as occupied
and active orbitals as well as a set of virtual DFS orbitals being the solutions of the DFS equations.
In our case, two different solutions of the SCF equations specify different
mutually non-orthogonal sets of the occupied DF orbitals. To construct a unified
basis set of the one-electron functions, we proceed as follows. The radial wave
functions of the occupied atomic shells except for the $4f$ shell are defined by
\begin{equation}
\left \{
\begin{array}{lll}
P_{a}(r) &=& 0.5 \, [P^{\rm (in)}_{a}(r)+P^{\rm (out)}_{a}(r)]
\\[1mm]
Q_{a}(r) &=& 0.5 \, [Q^{\rm (in)}_{a}(r)+Q^{\rm (out)}_{a}(r)] \,.
\end{array} \right .
\end{equation}
where the index $a$, as before, enumerates the atomic shells, 
$P^{\rm (in)}_{a}(r)$ and $Q^{\rm (in)}_{a}(r)$  are the large and small
components of the radial orbitals being the solutions of the DF equations
when the $4f$ orbital is localized in the inner well, and
$P^{\rm (out)}_{a}$ and $Q^{\rm (out)}_{a}$ are the same quantities for the case when the $4f$ orbital is in the outer well.  
Then, both $4f_{5/2}$ orbitals and a set of the virtual 
DFS orbitals are added to this unified basis, 
followed by an orthonormalization.
In our calculations,
besides the occupied DF orbitals, we have included $5d$ as the active orbital
and $7s$--$10s$, $6p$--$10p$, $6d$--$9d$, $5f$--$9f$, and $5g$--$6g$ as the virtual DFS ones.

\begin{table}
\caption{\label{tab_CI} 
The CI-DFS calculations of La, $Z=57$ ([Xe]$6s^2 4f_{5/2}^1$). 
$E_{\rm CI}$ is the total CI energies, $q_{4f_{5/2}}$ are the atomic
populations of the $4f_{5/2}$ shell. The labels ``in'' and ``out'' indicate two dual states. All the values are given in atomic units.}
%%%%%%%%%%%%%%%%%%%%%%%%%%%%%%%%%%%%%%%%%%%%%%%%%%%%%%%%%%%%%%%%%%%%%%%%%
\begin{ruledtabular}
\begin{tabular}{ccccc}
&  $-E_{\rm CI}$  & $q_{4f_{5/2}}^{\rm in}$ & 
$q_{4f_{5/2}}^{\rm out}$ & $q_{6s}$ \\
\hline
 ``in state'' &  8486.7738  &  0.98766  &  0.02583  &  1.83645 \\
 ``out state'' &  8486.5441  &  0.00006  &  0.96557  &  1.66580 \\
\end{tabular}
\end{ruledtabular}
\end{table}
To interpret our results, we employed an atomic-population analysis based
on the use of a one-particle density matrix $\rho$ in the atomic basis, 
or, in other words, the first-order reduced density matrix:
\begin{equation}
 \rho_{ij}=\langle \Psi \mid \hat{a}_i^+ \, \hat{a}_j \mid \Psi \rangle\,, \qquad
 q_a=\sum_{i \in a} \rho_{ii} \,,
\end{equation}
where $\Psi$ is the many-electron wave function, $ \hat{a}_i^+$ and $ \hat{a}_j$ are the creation and annihilation operators of the $i$-th and $j$-th electron, respectively, $q_a$ is the population
of the shell $a$, and the index $i$ enumerates all atomic orbitals of the shell $a$.
Based on the population analysis, we have identified
two eigenvectors of the CI matrix that give the configuration closest to
[Xe]$6s^2 4f_{5/2}^1$. One of these states, where the $4f$ electron is localized in the inner well, we denote as the ``in state'', and the other one, where the $4f$ electron
is localized in the outer well, is referred to as the ``out state''. As can be seen from 
Table~\ref{tab_CI}, the ``in state'' and "out state" interact weakly.
This can be explained by the small overlap of the $4f$ orbitals localized in
the inner and outer wells.  We also note that in this case the ``in state''
is energetically more favorable.
% %%%%%%%%%%%%%%%%%%%%%%%%%%%%%%%%%%%%%%%%%%%%%%%%%%%%%%%%%%%%%%%%%%%%%
\section{\label{sec:concl} Conclusions}
%%%%%%%%%%%%%%%%%%%%%%%%%%%%%%%%%%%%%%%%%%%%%%%%%%%%%%%%%%%%%%%%%%%%%
%
In the present work, it was found that the effective $5g$ or $6f$ radial
potentials for the eighth-period elements of the periodic table with the atomic
numbers $Z=125$, $Z=124$, $Z=134$, and $Z=148$ are double-well.  As a consequence, the
orbital collapse is observed in these elements.  For example, it has been
shown that, when the atomic term $J$ is changed in atoms with $Z=124,125$,
the wave function of the $5g$ electron, localized in the wide and shallow
outer well, shrinks strongly and turns out to be localized in the inner well.
As a result, the average radius of the $5g$ orbital decreases by a factor
of almost 40, and the binding energy of the $5g$ electron increases
by a factor of more than 25. It is shown, that the state of the $5g$ electron
in the outer well can be described with the high accuracy by the hydrogen wave
function. The state of the neutral atom with one electron in the outer
well can be interpreted as the motion of an electron in the Coulomb field
of a singly charged positive ion. The orbital-collapse effect for the $5g$
electrons is manifested more strongly than for the $4f$ and $6f$ electrons
because of the larger magnitude of the centrifugal term.

In this paper, we have confirmed the coexistence of the dual SCF solutions of
the same DF equations for La and Eu atoms observed earlier in 
Ref.~\cite{Band_1980}. In one of these solutions, the $4f$ electron is
localized in the inner well, whereas in the other solution it is localized in the outer well.
The similar dual states were found for the $5g$ electrons in the atoms with 
$Z=124, 125, 134$ and for the $6f$ electron in the atom with $Z=148$.

In order to verify the coexistence of the dual states in the many-electron
approach, on the example of the [Xe]$6s^2 4f_{5/2}^1$ configuration
of the La atom, we have performed the CI calculations 
including both states into the many-electron CSF basis.
Using the atomic-population analysis, it was found that
both dual states remain sufficiently stable when the
configuration interaction is taken into account. The final answer
on the question whether these dual states are actually physically
observable requires large-scale multi-configuration calculations.

In all examples of the orbital collapse and coexistence of the dual solutions,
we considered the configurations with one electron in the $4f_{5/2}$, $4f_{7/2}$,
$5g_{7/2}$, $5g_{9/2}$,  and $6f_{7/2}$ shells. This does not mean
that these effects cannot be observed for a larger number of electrons in these shells. However, in these cases, it is necessary to discard the central-field approximation,
according to which the radial functions of the different orbitals of 
the same shell
must be identical, and a configuration with more than one electron in the
outer well is energetically unfavorable.

\section{Acknowledgements}
The work is supported by the Ministry of Science and Higher Education of
the Russian Federation within the~Grant~No.~075-10-2020-117. The numerical
calculations are carried out using computing resources of the HybriLIT
heterogeneous computing platform (LIT, JINR)
(http://hlit.jinr.ru [hlit.jinr.ru]).
%
%%%%%%%%%%%%%%%%%%%%%%%%%%%%%%%%%%%%%%%%%%%%%%%%%%%%%%%%%%%%%%%%%%%%%%%%%%%%%
\bibliography{orbital_collapse}

\begin{thebibliography}{35}%
\makeatletter
\providecommand \@ifxundefined [1]{%
 \@ifx{#1\undefined}
}%
\providecommand \@ifnum [1]{%
 \ifnum #1\expandafter \@firstoftwo
 \else \expandafter \@secondoftwo
 \fi
}%
\providecommand \@ifx [1]{%
 \ifx #1\expandafter \@firstoftwo
 \else \expandafter \@secondoftwo
 \fi
}%
\providecommand \natexlab [1]{#1}%
\providecommand \enquote  [1]{``#1''}%
\providecommand \bibnamefont  [1]{#1}%
\providecommand \bibfnamefont [1]{#1}%
\providecommand \citenamefont [1]{#1}%
\providecommand \href@noop [0]{\@secondoftwo}%
\providecommand \href [0]{\begingroup \@sanitize@url \@href}%
\providecommand \@href[1]{\@@startlink{#1}\@@href}%
\providecommand \@@href[1]{\endgroup#1\@@endlink}%
\providecommand \@sanitize@url [0]{\catcode `\\12\catcode `\$12\catcode `\&12\catcode `\#12\catcode `\^12\catcode `\_12\catcode `\%12\relax}%
\providecommand \@@startlink[1]{}%
\providecommand \@@endlink[0]{}%
\providecommand \url  [0]{\begingroup\@sanitize@url \@url }%
\providecommand \@url [1]{\endgroup\@href {#1}{\urlprefix }}%
\providecommand \urlprefix  [0]{URL }%
\providecommand \Eprint [0]{\href }%
\providecommand \doibase [0]{https://doi.org/}%
\providecommand \selectlanguage [0]{\@gobble}%
\providecommand \bibinfo  [0]{\@secondoftwo}%
\providecommand \bibfield  [0]{\@secondoftwo}%
\providecommand \translation [1]{[#1]}%
\providecommand \BibitemOpen [0]{}%
\providecommand \bibitemStop [0]{}%
\providecommand \bibitemNoStop [0]{.\EOS\space}%
\providecommand \EOS [0]{\spacefactor3000\relax}%
\providecommand \BibitemShut  [1]{\csname bibitem#1\endcsname}%
\let\auto@bib@innerbib\@empty
%</preamble>
\bibitem [{\citenamefont {Fermi}(1928)}]{Fermi_1928}%
  \BibitemOpen
  \bibfield  {author} {\bibinfo {author} {\bibfnamefont {E.}~\bibnamefont {Fermi}},\ }\bibfield  {title} {\bibinfo {title} {{\"U}ber die anwendung der statistischen methode auf die probleme des atombaues},\ }in\ \href@noop {} {\emph {\bibinfo {booktitle} {Quantentheorie und Chemie}}},\ \bibinfo {editor} {edited by\ \bibinfo {editor} {\bibfnamefont {H.}~\bibnamefont {Falkenhagen}}}\ (\bibinfo  {publisher} {Leipziger Vortr{\"a}ge},\ \bibinfo {address} {Leipzig},\ \bibinfo {year} {1928})\ pp.\ \bibinfo {pages} {95--111}\BibitemShut {NoStop}%
\bibitem [{\citenamefont {Goeppert~Mayer}(1941)}]{Mayer_1941}%
  \BibitemOpen
  \bibfield  {author} {\bibinfo {author} {\bibfnamefont {M.}~\bibnamefont {Goeppert~Mayer}},\ }\bibfield  {title} {\bibinfo {title} {Rare-earth and transuranic elements},\ }\href {https://doi.org/10.1103/PhysRev.60.184} {\bibfield  {journal} {\bibinfo  {journal} {Phys. Rev.}\ }\textbf {\bibinfo {volume} {60}},\ \bibinfo {pages} {184} (\bibinfo {year} {1941})}\BibitemShut {NoStop}%
\bibitem [{\citenamefont {Cowan}(1968)}]{Cowan_1968}%
  \BibitemOpen
  \bibfield  {author} {\bibinfo {author} {\bibfnamefont {R.~D.}\ \bibnamefont {Cowan}},\ }\bibfield  {title} {\bibinfo {title} {Theoretical study of $p^{m} - p^{m -1}l$ spectra},\ }\href {https://doi.org/10.1364/JOSA.58.000924} {\bibfield  {journal} {\bibinfo  {journal} {J. Opt. Soc. Am.}\ }\textbf {\bibinfo {volume} {58}},\ \bibinfo {pages} {924} (\bibinfo {year} {1968})}\BibitemShut {NoStop}%
\bibitem [{\citenamefont {Griffin}\ \emph {et~al.}(1969)\citenamefont {Griffin}, \citenamefont {Andrew},\ and\ \citenamefont {Cowan}}]{Griffen_1969}%
  \BibitemOpen
  \bibfield  {author} {\bibinfo {author} {\bibfnamefont {D.~C.}\ \bibnamefont {Griffin}}, \bibinfo {author} {\bibfnamefont {K.~L.}\ \bibnamefont {Andrew}},\ and\ \bibinfo {author} {\bibfnamefont {R.~D.}\ \bibnamefont {Cowan}},\ }\bibfield  {title} {\bibinfo {title} {Theoretical calculations of the $d$-, $f$-, and $g$-electron transition series},\ }\href {https://doi.org/10.1103/PhysRev.177.62} {\bibfield  {journal} {\bibinfo  {journal} {Phys. Rev.}\ }\textbf {\bibinfo {volume} {177}},\ \bibinfo {pages} {62} (\bibinfo {year} {1969})}\BibitemShut {NoStop}%
\bibitem [{\citenamefont {Cheng}\ and\ \citenamefont {Froese~Fischer}(1983)}]{Cheng_1983}%
  \BibitemOpen
  \bibfield  {author} {\bibinfo {author} {\bibfnamefont {K.~T.}\ \bibnamefont {Cheng}}\ and\ \bibinfo {author} {\bibfnamefont {C.}~\bibnamefont {Froese~Fischer}},\ }\bibfield  {title} {\bibinfo {title} {Collapse of the $4f$ orbital for {Xe}-like ions},\ }\href {https://doi.org/10.1103/PhysRevA.28.2811} {\bibfield  {journal} {\bibinfo  {journal} {Phys. Rev. A}\ }\textbf {\bibinfo {volume} {28}},\ \bibinfo {pages} {2811} (\bibinfo {year} {1983})}\BibitemShut {NoStop}%
\bibitem [{\citenamefont {Migdalek}\ and\ \citenamefont {Baylis}(1984)}]{Migdalek_1984}%
  \BibitemOpen
  \bibfield  {author} {\bibinfo {author} {\bibfnamefont {J.}~\bibnamefont {Migdalek}}\ and\ \bibinfo {author} {\bibfnamefont {W.~E.}\ \bibnamefont {Baylis}},\ }\bibfield  {title} {\bibinfo {title} {Valence-core electron exchange interaction and the collapse of $4f$ and $5d$ orbitals in the cesium isoelectronic sequence},\ }\href {https://doi.org/10.1103/PhysRevA.30.1603} {\bibfield  {journal} {\bibinfo  {journal} {Phys. Rev. A}\ }\textbf {\bibinfo {volume} {30}},\ \bibinfo {pages} {1603} (\bibinfo {year} {1984})}\BibitemShut {NoStop}%
\bibitem [{\citenamefont {Migdalek}\ and\ \citenamefont {Siegel}(2000)}]{Migdalek_2000}%
  \BibitemOpen
  \bibfield  {author} {\bibinfo {author} {\bibfnamefont {J.}~\bibnamefont {Migdalek}}\ and\ \bibinfo {author} {\bibfnamefont {W.}~\bibnamefont {Siegel}},\ }\bibfield  {title} {\bibinfo {title} {Collapse of $d$ and $f$ orbitals in the isoelectronic sequence of singly ionized ytterbium},\ }\href {https://doi.org/10.1103/PhysRevA.61.062502} {\bibfield  {journal} {\bibinfo  {journal} {Phys. Rev. A}\ }\textbf {\bibinfo {volume} {61}},\ \bibinfo {pages} {062502} (\bibinfo {year} {2000})}\BibitemShut {NoStop}%
\bibitem [{\citenamefont {Lucatorto}\ \emph {et~al.}(1981)\citenamefont {Lucatorto}, \citenamefont {McIlrath}, \citenamefont {Sugar},\ and\ \citenamefont {Younger}}]{Lucatorto_1981}%
  \BibitemOpen
  \bibfield  {author} {\bibinfo {author} {\bibfnamefont {T.~J.}\ \bibnamefont {Lucatorto}}, \bibinfo {author} {\bibfnamefont {T.}~\bibnamefont {McIlrath}}, \bibinfo {author} {\bibfnamefont {J.}~\bibnamefont {Sugar}},\ and\ \bibinfo {author} {\bibfnamefont {S.~M.}\ \bibnamefont {Younger}},\ }\bibfield  {title} {\bibinfo {title} {Radical redistribution of the $4d$ oscillator strength observed in the photoabsorption of the {Ba}, {Ba}$^{+}$, and {Ba}$^{++}$ sequence},\ }\href {https://doi.org/10.1103/PhysRevLett.47.1124} {\bibfield  {journal} {\bibinfo  {journal} {Phys. Rev. Lett.}\ }\textbf {\bibinfo {volume} {47}},\ \bibinfo {pages} {1124} (\bibinfo {year} {1981})}\BibitemShut {NoStop}%
\bibitem [{\citenamefont {Connerade}(2000)}]{Connerade_2000}%
  \BibitemOpen
  \bibfield  {author} {\bibinfo {author} {\bibfnamefont {J.~P.}\ \bibnamefont {Connerade}},\ }\bibfield  {title} {\bibinfo {title} {Confined atoms: a new path towards controlled orbital collapse},\ }in\ \href@noop {} {\emph {\bibinfo {booktitle} {Trends in Atomic and Molecular Physics}}},\ \bibinfo {editor} {edited by\ \bibinfo {editor} {\bibfnamefont {K.~K.}\ \bibnamefont {Sud}}\ and\ \bibinfo {editor} {\bibfnamefont {U.~N.}\ \bibnamefont {Upadhyaya}}}\ (\bibinfo  {publisher} {Springer},\ \bibinfo {address} {Boston, MA},\ \bibinfo {year} {2000})\ pp.\ \bibinfo {pages} {235--249}\BibitemShut {NoStop}%
\bibitem [{\citenamefont {Connerade}(2020)}]{Connerade_2020}%
  \BibitemOpen
  \bibfield  {author} {\bibinfo {author} {\bibfnamefont {J.~P.}\ \bibnamefont {Connerade}},\ }\bibfield  {title} {\bibinfo {title} {Conﬁning and compressing the atom},\ }\href {https://doi.org/10.1140/epjd/e2020-10414-y} {\bibfield  {journal} {\bibinfo  {journal} {Eur. Phys. J. D}\ }\textbf {\bibinfo {volume} {74}},\ \bibinfo {pages} {211} (\bibinfo {year} {2020})}\BibitemShut {NoStop}%
\bibitem [{\citenamefont {Connerade}(1991)}]{Connerade_1991}%
  \BibitemOpen
  \bibfield  {author} {\bibinfo {author} {\bibfnamefont {J.~P.}\ \bibnamefont {Connerade}},\ }\bibfield  {title} {\bibinfo {title} {Orbital collapse in extended homologous sequences},\ }\href {https://doi.org/10.1088/0953-4075/24/5/001} {\bibfield  {journal} {\bibinfo  {journal} {J. Phys. B}\ }\textbf {\bibinfo {volume} {24}},\ \bibinfo {pages} {L109} (\bibinfo {year} {1991})}\BibitemShut {NoStop}%
\bibitem [{\citenamefont {Maiste}\ \emph {et~al.}(1980)\citenamefont {Maiste}, \citenamefont {Ruus}, \citenamefont {Kuchas}, \citenamefont {Karaziya},\ and\ \citenamefont {Elango}}]{Maiste_1980}%
  \BibitemOpen
  \bibfield  {author} {\bibinfo {author} {\bibfnamefont {A.~A.}\ \bibnamefont {Maiste}}, \bibinfo {author} {\bibfnamefont {R.~E.}\ \bibnamefont {Ruus}}, \bibinfo {author} {\bibfnamefont {S.~A.}\ \bibnamefont {Kuchas}}, \bibinfo {author} {\bibfnamefont {R.~I.}\ \bibnamefont {Karaziya}},\ and\ \bibinfo {author} {\bibfnamefont {M.~A.}\ \bibnamefont {Elango}},\ }\bibfield  {title} {\bibinfo {title} {Collapse of $4f$-electron in the configuration $3d^{9}4f$ in xenonlike ions},\ }\href@noop {} {\bibfield  {journal} {\bibinfo  {journal} {Zh. Eksp. Teor. Fiz.}\ }\textbf {\bibinfo {volume} {78}},\ \bibinfo {pages} {941} (\bibinfo {year} {1980})},\ \bibinfo {note} {[Sov. Phys. JETP \textbf{51}, 474 (1980)]}\BibitemShut {NoStop}%
\bibitem [{\citenamefont {Ruus}(1999)}]{Ruus_1999}%
  \BibitemOpen
  \bibfield  {author} {\bibinfo {author} {\bibfnamefont {R.}~\bibnamefont {Ruus}},\ }\emph {\bibinfo {title} {Collapse of $3d(4f)$ orbitals in $2p(3d)$ excited configurations and its effect on the X-Ray and electron spectra}},\ \href@noop {} {Ph.D. thesis},\ \bibinfo  {school} {University of Tartu, Estonia} (\bibinfo {year} {1999})\BibitemShut {NoStop}%
\bibitem [{\citenamefont {Tupitsyn}\ \emph {et~al.}(2023)\citenamefont {Tupitsyn}, \citenamefont {Savelyev}, \citenamefont {Kozhedub}, \citenamefont {Kaygorodov}, \citenamefont {Glazov}, \citenamefont {Dulaev}, \citenamefont {Malyshev},\ and\ \citenamefont {Shabaev}}]{Tupitsyn_2023}%
  \BibitemOpen
  \bibfield  {author} {\bibinfo {author} {\bibfnamefont {I.~I.}\ \bibnamefont {Tupitsyn}}, \bibinfo {author} {\bibfnamefont {I.~M.}\ \bibnamefont {Savelyev}}, \bibinfo {author} {\bibfnamefont {Y.~S.}\ \bibnamefont {Kozhedub}}, \bibinfo {author} {\bibfnamefont {M.~Y.}\ \bibnamefont {Kaygorodov}}, \bibinfo {author} {\bibfnamefont {D.~A.}\ \bibnamefont {Glazov}}, \bibinfo {author} {\bibfnamefont {N.~K.}\ \bibnamefont {Dulaev}}, \bibinfo {author} {\bibfnamefont {A.~V.}\ \bibnamefont {Malyshev}},\ and\ \bibinfo {author} {\bibfnamefont {V.~M.}\ \bibnamefont {Shabaev}},\ }\bibfield  {title} {\bibinfo {title} {Orbital collapse of $5g$-electrons in superheavy elements of the 8th period},\ }\href {https://doi.org/10.21883/OS.2023.07.56122.4747-22} {\bibfield  {journal} {\bibinfo  {journal} {Opt. Spektrosk.}\ }\textbf {\bibinfo {volume} {131}},\ \bibinfo {pages} {895} (\bibinfo {year} {2023})}\BibitemShut {NoStop}%
\bibitem [{\citenamefont {Connerade}(1978)}]{Connerade_1978}%
  \BibitemOpen
  \bibfield  {author} {\bibinfo {author} {\bibfnamefont {J.~P.}\ \bibnamefont {Connerade}},\ }\bibfield  {title} {\bibinfo {title} {The non-{R}ydberg spectroscopy of atoms},\ }\href {https://doi.org/10.1080/00107517808210893} {\bibfield  {journal} {\bibinfo  {journal} {Cont. Phys.}\ }\textbf {\bibinfo {volume} {19}},\ \bibinfo {pages} {415} (\bibinfo {year} {1978})}\BibitemShut {NoStop}%
\bibitem [{\citenamefont {Karaziya}(1981)}]{Karaziya_1981}%
  \BibitemOpen
  \bibfield  {author} {\bibinfo {author} {\bibfnamefont {R.~I.}\ \bibnamefont {Karaziya}},\ }\bibfield  {title} {\bibinfo {title} {Excited electron orbit collapse and atomic spectra},\ }\href {https://doi.org/10.1070/PU1981v024n09ABEH004823} {\bibfield  {journal} {\bibinfo  {journal} {Sov. Phys. Usp.}\ }\textbf {\bibinfo {volume} {24}},\ \bibinfo {pages} {775} (\bibinfo {year} {1981})}\BibitemShut {NoStop}%
\bibitem [{\citenamefont {Band}\ and\ \citenamefont {Fomichev}(1980)}]{Band_1980}%
  \BibitemOpen
  \bibfield  {author} {\bibinfo {author} {\bibfnamefont {I.~M.}\ \bibnamefont {Band}}\ and\ \bibinfo {author} {\bibfnamefont {I.~M.}\ \bibnamefont {Fomichev}},\ }\bibfield  {title} {\bibinfo {title} {Coexistence of “collapse” and “blow” states of the same atom in the rare-earth region},\ }\href {https://doi.org/10.1016/0375-9601(80)90106-1} {\bibfield  {journal} {\bibinfo  {journal} {Phys. Lett. A}\ }\textbf {\bibinfo {volume} {75}},\ \bibinfo {pages} {178} (\bibinfo {year} {1980})}\BibitemShut {NoStop}%
\bibitem [{\citenamefont {Band}\ \emph {et~al.}(1981)\citenamefont {Band}, \citenamefont {Fomichev},\ and\ \citenamefont {Trzhaskovskaya}}]{Band_1981}%
  \BibitemOpen
  \bibfield  {author} {\bibinfo {author} {\bibfnamefont {I.~M.}\ \bibnamefont {Band}}, \bibinfo {author} {\bibfnamefont {I.~M.}\ \bibnamefont {Fomichev}},\ and\ \bibinfo {author} {\bibfnamefont {M.~B.}\ \bibnamefont {Trzhaskovskaya}},\ }\bibfield  {title} {\bibinfo {title} {'{D}irac-{F}ock atoms' in the rare-earth region and the 4f wavefunction collapse phenomenon},\ }\href {https://doi.org/10.1088/0022-3700/14/7/008} {\bibfield  {journal} {\bibinfo  {journal} {J. Phys. B}\ }\textbf {\bibinfo {volume} {14}},\ \bibinfo {pages} {1103} (\bibinfo {year} {1981})}\BibitemShut {NoStop}%
\bibitem [{\citenamefont {Migdalek}\ and\ \citenamefont {Baylis}(1987)}]{Migdalek_1987}%
  \BibitemOpen
  \bibfield  {author} {\bibinfo {author} {\bibfnamefont {J.}~\bibnamefont {Migdalek}}\ and\ \bibinfo {author} {\bibfnamefont {W.~E.}\ \bibnamefont {Baylis}},\ }\bibfield  {title} {\bibinfo {title} {A multiconfiguration {D}irac-{F}ock study of the $6s^{2}$ $^{1}{S}_{0}$ -- $6s6p$ $^{3}{P}_{1}$, $^{1}{P}_{1}$ transitions in the {Yb} isoelectronic sequence},\ }\href {https://doi.org/10.1016/0022-4073(87)90054-9} {\bibfield  {journal} {\bibinfo  {journal} {J. Quant. Spectrosc. Radiat. Transf.}\ }\textbf {\bibinfo {volume} {37}},\ \bibinfo {pages} {521} (\bibinfo {year} {1987})}\BibitemShut {NoStop}%
\bibitem [{\citenamefont {Atkinson}(1964)}]{Atkinson_1964}%
  \BibitemOpen
  \bibinfo {editor} {\bibfnamefont {F.~V.}\ \bibnamefont {Atkinson}},\ ed.,\ \href@noop {} {\emph {\bibinfo {title} {Discrete and Continuous Boundary Problems}}}\ (\bibinfo  {publisher} {Academic {P}ress},\ \bibinfo {address} {New {Y}ork},\ \bibinfo {year} {1964})\BibitemShut {NoStop}%
\bibitem [{\citenamefont {Savelyev}\ \emph {et~al.}(2023)\citenamefont {Savelyev}, \citenamefont {Kaygorodov}, \citenamefont {Kozhedub}, \citenamefont {Malyshev}, \citenamefont {Tupitsyn},\ and\ \citenamefont {Shabaev}}]{Savelyev_2023}%
  \BibitemOpen
  \bibfield  {author} {\bibinfo {author} {\bibfnamefont {I.~M.}\ \bibnamefont {Savelyev}}, \bibinfo {author} {\bibfnamefont {M.~Y.}\ \bibnamefont {Kaygorodov}}, \bibinfo {author} {\bibfnamefont {Y.~S.}\ \bibnamefont {Kozhedub}}, \bibinfo {author} {\bibfnamefont {A.~V.}\ \bibnamefont {Malyshev}}, \bibinfo {author} {\bibfnamefont {I.~I.}\ \bibnamefont {Tupitsyn}},\ and\ \bibinfo {author} {\bibfnamefont {V.~M.}\ \bibnamefont {Shabaev}},\ }\bibfield  {title} {\bibinfo {title} {Ground state of superheavy elements with $120\leqslant {Z}\leqslant 170$: {S}ystematic study of the electron-correlation, {B}reit, and {QED} effects},\ }\href {https://doi.org/10.1103/PhysRevA.107.042803} {\bibfield  {journal} {\bibinfo  {journal} {Phys. Rev. A}\ }\textbf {\bibinfo {volume} {107}},\ \bibinfo {pages} {042803} (\bibinfo {year} {2023})}\BibitemShut {NoStop}%
\bibitem [{\citenamefont {Fricke}\ and\ \citenamefont {Soff}(1977)}]{Fricke_1977}%
  \BibitemOpen
  \bibfield  {author} {\bibinfo {author} {\bibfnamefont {B.}~\bibnamefont {Fricke}}\ and\ \bibinfo {author} {\bibfnamefont {G.}~\bibnamefont {Soff}},\ }\bibfield  {title} {\bibinfo {title} {{D}irac-{F}ock-{S}later calculations for the elements ${Z} = 100$, fermium, to ${Z} = 173$},\ }\href {https://doi.org/10.1016/0092-640X(77)90010-9} {\bibfield  {journal} {\bibinfo  {journal} {At. Data Nucl. Data Tables}\ }\textbf {\bibinfo {volume} {19}},\ \bibinfo {pages} {83} (\bibinfo {year} {1977})}\BibitemShut {NoStop}%
\bibitem [{\citenamefont {Nefedov}\ \emph {et~al.}(2006)\citenamefont {Nefedov}, \citenamefont {Trzhaskovskaya},\ and\ \citenamefont {Yarzhemskii}}]{Nefedov_2006}%
  \BibitemOpen
  \bibfield  {author} {\bibinfo {author} {\bibfnamefont {V.~I.}\ \bibnamefont {Nefedov}}, \bibinfo {author} {\bibfnamefont {M.~B.}\ \bibnamefont {Trzhaskovskaya}},\ and\ \bibinfo {author} {\bibfnamefont {V.~G.}\ \bibnamefont {Yarzhemskii}},\ }\bibfield  {title} {\bibinfo {title} {Electronic configurations and the periodic table for superheavy elements},\ }\href {https://doi.org/10.1134/S0012501606060029} {\bibfield  {journal} {\bibinfo  {journal} {Dokl. Phys. Chem.}\ }\textbf {\bibinfo {volume} {408}},\ \bibinfo {pages} {149} (\bibinfo {year} {2006})}\BibitemShut {NoStop}%
\bibitem [{\citenamefont {Smits}\ \emph {et~al.}(2023{\natexlab{a}})\citenamefont {Smits}, \citenamefont {Indelicato}, \citenamefont {Nazarewicz}, \citenamefont {Piibeleht},\ and\ \citenamefont {Schwerdtfeger}}]{Smits_2023a}%
  \BibitemOpen
  \bibfield  {author} {\bibinfo {author} {\bibfnamefont {O.~R.}\ \bibnamefont {Smits}}, \bibinfo {author} {\bibfnamefont {P.}~\bibnamefont {Indelicato}}, \bibinfo {author} {\bibfnamefont {W.}~\bibnamefont {Nazarewicz}}, \bibinfo {author} {\bibfnamefont {M.}~\bibnamefont {Piibeleht}},\ and\ \bibinfo {author} {\bibfnamefont {P.}~\bibnamefont {Schwerdtfeger}},\ }\bibfield  {title} {\bibinfo {title} {Pushing the limits of the periodic table — a review on atomic relativistic electronic structure theory and calculations for the superheavy elements},\ }\href {https://doi.org/10.1016/j.physrep.2023.09.004} {\bibfield  {journal} {\bibinfo  {journal} {Physics Reports}\ }\textbf {\bibinfo {volume} {1035}},\ \bibinfo {pages} {1} (\bibinfo {year} {2023}{\natexlab{a}})}\BibitemShut {NoStop}%
\bibitem [{\citenamefont {Smits}\ \emph {et~al.}(2023{\natexlab{b}})\citenamefont {Smits}, \citenamefont {D\"ullmann}, \citenamefont {Indelicato}, \citenamefont {Nazarewicz},\ and\ \citenamefont {Schwerdtfeger}}]{Smits_2023b}%
  \BibitemOpen
  \bibfield  {author} {\bibinfo {author} {\bibfnamefont {O.~R.}\ \bibnamefont {Smits}}, \bibinfo {author} {\bibfnamefont {C.~E.}\ \bibnamefont {D\"ullmann}}, \bibinfo {author} {\bibfnamefont {P.}~\bibnamefont {Indelicato}}, \bibinfo {author} {\bibfnamefont {W.}~\bibnamefont {Nazarewicz}},\ and\ \bibinfo {author} {\bibfnamefont {P.}~\bibnamefont {Schwerdtfeger}},\ }\bibfield  {title} {\bibinfo {title} {The quest for superheavy elements and the limit of the periodic table},\ }\href {https://doi.org/10.1038/s42254-023-00668-y} {\bibfield  {journal} {\bibinfo  {journal} {Nature Reviews Physics}\ } (\bibinfo {year} {2023}{\natexlab{b}})},\ \bibinfo {note} {https://doi.org/10.1038/s42254-023-00668-y}\BibitemShut {NoStop}%
\bibitem [{\citenamefont {Bratzev}\ \emph {et~al.}(1977)\citenamefont {Bratzev}, \citenamefont {Deyneka},\ and\ \citenamefont {Tupitsyn}}]{Bratzev_1977}%
  \BibitemOpen
  \bibfield  {author} {\bibinfo {author} {\bibfnamefont {V.~F.}\ \bibnamefont {Bratzev}}, \bibinfo {author} {\bibfnamefont {G.~B.}\ \bibnamefont {Deyneka}},\ and\ \bibinfo {author} {\bibfnamefont {I.~I.}\ \bibnamefont {Tupitsyn}},\ }\bibfield  {title} {\bibinfo {title} {Application of {H}artree-{F}ock method to calculation of relativistic atomic wave functions},\ }\href@noop {} {\bibfield  {journal} {\bibinfo  {journal} {Bull. Acad. Sci. USSR, Phys. Ser.}\ }\textbf {\bibinfo {volume} {41}},\ \bibinfo {pages} {173} (\bibinfo {year} {1977})}\BibitemShut {NoStop}%
\bibitem [{\citenamefont {Grant}(1970)}]{Grant_1970}%
  \BibitemOpen
  \bibfield  {author} {\bibinfo {author} {\bibfnamefont {I.~P.}\ \bibnamefont {Grant}},\ }\bibfield  {title} {\bibinfo {title} {Relativistic calculation of atomic structures},\ }\href {https://doi.org/10.1080/00018737000101191} {\bibfield  {journal} {\bibinfo  {journal} {Adv. Phys.}\ }\textbf {\bibinfo {volume} {19}},\ \bibinfo {pages} {747} (\bibinfo {year} {1970})}\BibitemShut {NoStop}%
\bibitem [{\citenamefont {Tupitsyn}\ \emph {et~al.}(2018)\citenamefont {Tupitsyn}, \citenamefont {Zubova}, \citenamefont {Shabaev}, \citenamefont {Plunien},\ and\ \citenamefont {Stöhlker}}]{Tupitsyn_2018}%
  \BibitemOpen
  \bibfield  {author} {\bibinfo {author} {\bibfnamefont {I.~I.}\ \bibnamefont {Tupitsyn}}, \bibinfo {author} {\bibfnamefont {N.~A.}\ \bibnamefont {Zubova}}, \bibinfo {author} {\bibfnamefont {V.~M.}\ \bibnamefont {Shabaev}}, \bibinfo {author} {\bibfnamefont {G.}~\bibnamefont {Plunien}},\ and\ \bibinfo {author} {\bibfnamefont {T.}~\bibnamefont {Stöhlker}},\ }\bibfield  {title} {\bibinfo {title} {Relativistic calculations of x-ray transition energies and isotope shifts in heavy atoms},\ }\href {https://doi.org/10.1103/PhysRevA.98.022517} {\bibfield  {journal} {\bibinfo  {journal} {Phys. Rev. A}\ }\textbf {\bibinfo {volume} {98}},\ \bibinfo {pages} {022517} (\bibinfo {year} {2018})}\BibitemShut {NoStop}%
\bibitem [{\citenamefont {Tupitsyn}\ and\ \citenamefont {Loginov}(2003)}]{Tupitsyn_2003A}%
  \BibitemOpen
  \bibfield  {author} {\bibinfo {author} {\bibfnamefont {I.~I.}\ \bibnamefont {Tupitsyn}}\ and\ \bibinfo {author} {\bibfnamefont {A.~V.}\ \bibnamefont {Loginov}},\ }\bibfield  {title} {\bibinfo {title} {Use of {S}turmian expansions in calculations of the hyperfine structure of atomic spectra},\ }\href {https://doi.org/10.1134/1.1563671} {\bibfield  {journal} {\bibinfo  {journal} {Opt. Spectrosc.}\ }\textbf {\bibinfo {volume} {94}},\ \bibinfo {pages} {319} (\bibinfo {year} {2003})}\BibitemShut {NoStop}%
\bibitem [{\citenamefont {Tupitsyn}\ \emph {et~al.}(2003)\citenamefont {Tupitsyn}, \citenamefont {Shabaev}, \citenamefont {Crespo L\'opez-Urrutia}, \citenamefont {Dragani\'c}, \citenamefont {Orts},\ and\ \citenamefont {Ullrich}}]{Tupitsyn_2003B}%
  \BibitemOpen
  \bibfield  {author} {\bibinfo {author} {\bibfnamefont {I.~I.}\ \bibnamefont {Tupitsyn}}, \bibinfo {author} {\bibfnamefont {V.~M.}\ \bibnamefont {Shabaev}}, \bibinfo {author} {\bibfnamefont {J.~R.}\ \bibnamefont {Crespo L\'opez-Urrutia}}, \bibinfo {author} {\bibfnamefont {I.}~\bibnamefont {Dragani\'c}}, \bibinfo {author} {\bibfnamefont {R.~S.}\ \bibnamefont {Orts}},\ and\ \bibinfo {author} {\bibfnamefont {J.}~\bibnamefont {Ullrich}},\ }\bibfield  {title} {\bibinfo {title} {Relativistic calculations of isotope shifts in highly charged ions},\ }\href {https://doi.org/10.1103/PhysRevA.68.022511} {\bibfield  {journal} {\bibinfo  {journal} {Phys. Rev. A}\ }\textbf {\bibinfo {volume} {68}},\ \bibinfo {pages} {022511} (\bibinfo {year} {2003})}\BibitemShut {NoStop}%
\bibitem [{\citenamefont {Pieper}\ and\ \citenamefont {Greiner}(1969)}]{Pieper_1969}%
  \BibitemOpen
  \bibfield  {author} {\bibinfo {author} {\bibfnamefont {W.}~\bibnamefont {Pieper}}\ and\ \bibinfo {author} {\bibfnamefont {W.}~\bibnamefont {Greiner}},\ }\bibfield  {title} {\bibinfo {title} {Interior electron shells in superheavy nuclei},\ }\href {https://doi.org/10.1007/BF01670014} {\bibfield  {journal} {\bibinfo  {journal} {Z. Phys.}\ }\textbf {\bibinfo {volume} {218}},\ \bibinfo {pages} {327} (\bibinfo {year} {1969})}\BibitemShut {NoStop}%
\bibitem [{\citenamefont {Angeli}\ and\ \citenamefont {Marinova}(2013)}]{Angeli_2013}%
  \BibitemOpen
  \bibfield  {author} {\bibinfo {author} {\bibfnamefont {I.}~\bibnamefont {Angeli}}\ and\ \bibinfo {author} {\bibfnamefont {K.~P.}\ \bibnamefont {Marinova}},\ }\bibfield  {title} {\bibinfo {title} {Table of experimental nuclear ground state charge radii: {An} update},\ }\href {https://doi.org/10.1016/j.adt.2011.12.006} {\bibfield  {journal} {\bibinfo  {journal} {At. Data Nucl. Data Tables}\ }\textbf {\bibinfo {volume} {99}},\ \bibinfo {pages} {69} (\bibinfo {year} {2013})}\BibitemShut {NoStop}%
\bibitem [{\citenamefont {G{\'a}sp{\'a}r}(1952)}]{Gaspar_1952}%
  \BibitemOpen
  \bibfield  {author} {\bibinfo {author} {\bibfnamefont {R.}~\bibnamefont {G{\'a}sp{\'a}r}},\ }\bibfield  {title} {\bibinfo {title} {An analytical method for the approximate determination of the eigenfunctions and energies of electrons in atoms},\ }\href {https://doi.org/10.1063/1.1700328} {\bibfield  {journal} {\bibinfo  {journal} {J. Chem. Phys.}\ }\textbf {\bibinfo {volume} {20}},\ \bibinfo {pages} {1863} (\bibinfo {year} {1952})}\BibitemShut {NoStop}%
\bibitem [{\citenamefont {Green}(1973)}]{Green_1973}%
  \BibitemOpen
  \bibfield  {author} {\bibinfo {author} {\bibfnamefont {A.~E.~S.}\ \bibnamefont {Green}},\ }\bibfield  {title} {\bibinfo {title} {An analytic independent particle model for atoms: {I.} {I}nitial studies},\ }\href {https://doi.org/10.1016/S0065-3276(08)60563-8} {\bibfield  {journal} {\bibinfo  {journal} {Adv. Quant. Chem.}\ }\textbf {\bibinfo {volume} {7}},\ \bibinfo {pages} {221} (\bibinfo {year} {1973})}\BibitemShut {NoStop}%
\bibitem [{\citenamefont {Shabaev}\ \emph {et~al.}(2005)\citenamefont {Shabaev}, \citenamefont {Tupitsyn}, \citenamefont {Pachucki}, \citenamefont {Plunien},\ and\ \citenamefont {Yerokhin}}]{Shabaev_2005}%
  \BibitemOpen
  \bibfield  {author} {\bibinfo {author} {\bibfnamefont {V.~M.}\ \bibnamefont {Shabaev}}, \bibinfo {author} {\bibfnamefont {I.~I.}\ \bibnamefont {Tupitsyn}}, \bibinfo {author} {\bibfnamefont {K.}~\bibnamefont {Pachucki}}, \bibinfo {author} {\bibfnamefont {G.}~\bibnamefont {Plunien}},\ and\ \bibinfo {author} {\bibfnamefont {V.~A.}\ \bibnamefont {Yerokhin}},\ }\bibfield  {title} {\bibinfo {title} {Radiative and correlation effects on the parity-nonconserving transition amplitude in heavy alkali-metal atoms},\ }\href {https://doi.org/10.1103/PhysRevA.72.062105} {\bibfield  {journal} {\bibinfo  {journal} {Phys. Rev. A}\ }\textbf {\bibinfo {volume} {72}},\ \bibinfo {pages} {062105} (\bibinfo {year} {2005})}\BibitemShut {NoStop}%
\end{thebibliography}%
%%%%%%%%%%%%%%%%%%%%%%%%%%%%%%%%%%%%%%%%%%%%%%%%%%%%%%%%%%%%%%%%%%%%%%%%%%%%%
\end {document}